\documentstyle[12pt]{article}

\def\beq{\begin{equation}}
\def\eeq{\end{equation}}
\def\bea{\begin{eqnarray}}
\def\eea{\end{eqnarray}}

\def\ba{\begin{array}}
\def\ea{\end{array}}

\def\,{\"{U}}
\def\6{\.{I}}

\begin{document}

\title{A new approach to the exact solutions of the effective mass Schr\"{o}dinger equation}
\author{\vspace{1cm}
{Cevdet Tezcan $^1$\thanks{Corresponding Author:
sever@metu.edu.tr}, Ramazan Sever $^2$, \"{O}zlem Ye\c{s}ilta\c{s}
$^3$}
         \\
{\small \sl  $^1$Faculty of Engineering, Ba\c{s}kent University,
Ba\~{g}l{\i}ca Campus, Ankara, Turkey }
\\{\small \sl $^2$Middle East Technical University,Department of
Physics, 06531 Ankara, Turkey }
\\ {\small \sl $^3$ Gazi University, Faculty of Arts and Sciences, Department of Physics,
06500, Ankara,Turkey}}
\date{\today}
\maketitle
\begin{abstract}
\noindent Effective mass Schr\"{o}dinger equation is solved exactly
for a given potential. Nikiforov-Uvarov method is used to obtain
energy eigenvalues and the corresponding wave functions. A free
parameter is used in the transformation of the wave function. The
effective mass Schr\"{o}dinger equation is also solved for the Morse
potential transforming to the constant mass Schr\"{o}dinger equation
for a potential. One can also get solution of the effective mass
Schr\"{o}dinger equation starting from the constant mass
Schr\"{o}dinger equation.
\\
{\small \sl \noindent PACS numbers: 03.65.-w; 03.65.Ge; 12.39.Fd \\[0.2cm]
}
\end{abstract}
\baselineskip 0.9cm
\newpage
\section{Introduction}

Quantum mechanical systems with  spatially dependent effective mass
(PDM) has been extensively used in different branch of physics.
Special PDM applications can be found in the fields of
microstructures such as semiconductors [1], quantum dots [2], helium
clusters [3], quantum liquids [4]. Content of the approach requires
non-constant mass that depends on position such that the mass and
the momentum operator no longer commute  in PDM Schr\"{o}dinger
equation [5]. Dekar et al. solved the one dimensional
Schr\"{o}dinger equation with smooth potential and mass step [6].
Fruitful applications have been increased in quantum mechanical
problems and various approaches are used in the atomic and nuclear
physics and other fields of the physics. Supersymmetric (SUSY)
method is very useful technique for exactly solvable potentials that
extended to PDM [7]. Several authors have investigated the exact
solution of Schr\"{o}dinger equation with position dependent mass
using SUSY techniques [8,9,10]. so(2,1), su(1,1) Lie algebras and
quadratic algebra approach for PDM Schr\"{o}dinger equation in two
dimensions were used as generating algebra as a potential algebra to
obtain exact solutions of the effective mass wave equation
[11,12,13]. Exact solutions of Schr\"{o}dinger equation in D
dimensions, quantum well problem includes PDM approach [14,15] and
the point canonical transformations (PCT) are other studies and
approaches providing exact solution of energy eigenvalues and
corresponding eigenfunctions [16,17,18,19].

Nikiforov-Uvarov approach which received much interest in recent
years, has been introduced for solving Schr\"{o}dinger equation,
Klein-Gordon, Dirac and Salpeter equations [20,21,22,23,24,25].

In this work, the PDM Schr\"{o}dinger equation is solved for a given
potential by using the Nikivori-Uvarov method. Energy eigenvalues
and the corresponding wave functions are calculated. Then  as an
application the PDM Schr\"{o}dinger equation is solved for the Morse
potential by transforming into a constant mass Schr\"{o}dinger
equation for a well known potential. It is also shown that solutions
of the PDM Schr\"{o}dinger equation can be obtained starting from
the constant mass Schr\"{o}dinger equation can be obtained for the
Morse potential.

The contents of the paper is as follows: in section II, we
introduce PDM approach and Nikiforov-Uvarov method. The next
section involves applications to Morse potential. Solutions
obtained with mass dependent parameters are given in section IV.
Results are discussed in section V.

\section{Method}

\noindent We write the one-dimensional effective mass Hamiltonian
of the SE as $\left[26\right]$

\begin{equation}
H_{eff}=-\frac{d}{dx}\left(\frac{1}{m(x)}\frac{d}{dx}\right)+V_{eff}(x)
\end{equation}
where $V_{eff}$ has the form

\begin{equation}
V_{eff}=V(x)+\frac{1}{2}(\beta+1)\frac{m^{''}}{m^{2}}-\left[\alpha(\alpha+\beta+1)+\beta+1\right]\frac{{m^{'}}^{2}}{m^{3}}
\end{equation}
with $\alpha$, $\beta$ are ambiguity parameters. Primes stand for
the derivatives with respect to $x$. Thus the SE takes the form

\begin{equation}
\left(-\frac{1}{m}\frac{d^{2}}{dx^{2}}+\frac{m^{\prime}}{m}\frac{d}{dx}+V_{eff}-\varepsilon\right)\varphi(x)=0
\end{equation}
We apply the following transformation

\begin{equation}
\varphi=m^{\eta}(x)\;\psi(x)
\end{equation}
Hence, the SE takes the form

\begin{equation}
\left\{-\frac{1}{m}\left[\frac{d^{2}}{dx^{2}}+(2\eta-1)\frac{{m^{'}}}{m}\frac{d}{dx}+
\eta\left((\eta-2)\left(\frac{m^{'}}{m}\right)^{2}+\frac{m^{''}}{m}\right)\right]+(V_{eff}-\varepsilon)
\right\}\psi=0
\end{equation}
For the case of $\eta=1/2$, this equation turns to Eq.(4) in
Ref.[26]. Now we assume that

\begin{eqnarray}
  m(x) &=& e^{-2\lambda x} \\
  V(x) &=& V_{0}e^{2\lambda x}-B(2A+1)e^{\lambda x}
\end{eqnarray}
Substituting these relations into Eq.(5), we get

\begin{equation}
-[\psi^{''}+2\lambda (2\eta-1)\psi^{'}+4\eta
\lambda^{2}(\eta-1)\psi]+ [V_{0}-B(2A+1)e^{-\lambda x}-\varepsilon
e^{-2\lambda x}+ 2(\beta+1)\lambda^{2}-4A^{*}\lambda^{2}]\psi=0
\end{equation}
where

\begin{equation}
A^{*}=\alpha(\alpha+\beta+1)+\beta+1
\end{equation}
The coordinate transformation $s=e^{-\lambda x}$ leads to

\begin{equation}
\frac{d^{2}\psi}{ds^{2}}+(3-4\eta)\frac{1}{s}\frac{d\psi}{ds}+
\frac{1}{s^{2}}\left[\frac{\varepsilon}{\lambda^{2}}s^{2}+\frac{1}{\lambda^{2}}B(2A+1)s-
\frac{V_{0}}{\lambda^{2}}-2(\beta+1)+4A^{*}+4\eta(\eta-1)\right]\psi=0.
\end{equation}
For simplicity, let us define

\begin{eqnarray}
  \xi_{1} &=& -\frac{\varepsilon}{\lambda^{2}} \\
  \xi_{2} &=& -\frac{1}{\lambda^{2}}B(2A+1) \\
  -\xi_{3} &=& \frac{V_{0}}{\lambda^{2}}+2(\beta+1)-4A^{*}-4\eta(\eta+1)
\end{eqnarray}
Thus, Eq.(10) has the form

\begin{equation}
\frac{d^{2}\psi}{ds^{2}}+(3-4\eta)\frac{1}{s}\frac{d\psi}{ds}+\frac{1}{s^{2}}(-\xi_{1}s^{2}-\xi_{2}s+\xi_{3})\psi=0
\end{equation}
Now, we apply the NU method starting from its standard form

\begin{equation}
\psi^{''}_{n}(s)+\frac{\tilde{\tau}(s)}{\sigma(s)}\psi^{'}(s)+
\frac{\tilde{\sigma}(s)}{\sigma^{2}(s)}\psi_{n}(s)=0.
\end{equation}
Comparing Eqs.(14) and (15), we obtain

\begin{equation}
\sigma=s,  \tilde{\tau}(s)=3-4\eta,
\tilde{\sigma}(s)=-\xi_{1}s^{2}-\xi_{2}s+\xi_{3}
\end{equation}
where $\sigma(s)$ and $\tilde{\sigma}(s)$ are polynomials at most
second degree and $\tilde{\tau}(s)$ is a first-degree polynomial.
In the NU method, the function $\pi$ and the parameter $\lambda$
are defined as

\begin{equation}
\pi(s)=\frac{\sigma^{'}-\tau(s)}{2}\pm
\sqrt{\left(\frac{\sigma^{'}-\tau(s)}{2}\right)^{2}-\tilde{\sigma}(s)+k\sigma(s)}
\end{equation}
and

\begin{equation}
\lambda=k+\pi^{'}
\end{equation}
To find a physical solution, the expression in  the square root must
be square of a polynomial. Then, a new eigenvalue equation for the
SE becomes

\begin{equation}
\lambda=\lambda_{n}=-n\tau^{'}-\frac{n(n-1)}{2}\sigma^{''}(s),
(n=0,1,2,...)
\end{equation}%
where

\begin{equation}
\tau(s)=\tilde{\tau}(s)+2\pi(s)
\end{equation}%
and it should have a negative derivative [20]. A family of
particular solutions for a given $\lambda$ has hypergeometric type
of degree. Thus, $\lambda=0$ will corresponds to energy eigenvalue
of the ground state, i.e. $n=0$. The wave function is obtained as
a multiple of two independent parts

\begin{equation}
\psi(s)=\phi(s)y(s)
\end{equation}%
where $y(s)$ is the hypergeometric type function written with a
weight function $\rho$ as

\begin{equation}
y_{n}(s)=\frac{B_{n}}{\rho(s)}\frac{d^{n}}{ds}[\sigma^{n}(s)\rho(s)]
\end{equation}%
where $\rho(s)$ must satisfy the condition [20]

\begin{equation}
(\sigma \rho)^{'}=\tau \rho
\end{equation}%
The other part is defined as a logarithmic derivative

\begin{equation}
\frac{\phi^{'}(s)}{\phi(s)}=\frac{\pi(s)}{\sigma(s)}
\end{equation}%

\section{Calculations}
\subsection{Solutions of Eq.(14) with the Nikiforov-Uvarov method}

Substituting $\sigma(s)$, $\tilde{\sigma}$ and $\tilde{\tau}(s)$
into Eq.(17), we obtain $\pi$ function as

\begin{equation}
\pi=2\eta-1\pm \sqrt{\xi_{1}s^{2}-2Ds+(2\eta-1)^{2}-\xi_{3}}
\end{equation}
Due to NU method, the expression in the square root is taken as
the square of a polynomial. Then, one gets the possible functions
for each root $k$ as

\begin{equation}
\pi=2\eta-1\pm
\end{equation}

\begin{equation}
\left\{%
\begin{array}{ll}
     \sqrt{\xi_{1}s^{2}+2Ds+(2\eta-1)^{2}-\xi_{3}}, & \hbox{$k_{1}=2\sqrt{\xi_{1}\left[(2\eta-1)^{2}-\xi_{3}\right]}-\xi_{2}$;} \\
     \sqrt{\xi_{1}s^{2}-2Ds+(2\eta-1)^{2}-\xi_{3}}, & \hbox{$k_{2}=-2\sqrt{\xi_{1}\left[(2\eta-1)^{2}-\xi_{3}\right]}-\xi_{2}$.} \\
\end{array}%
\right.
\end{equation}
where $D^{2}=\xi_{1}[(2\eta-1)^{2}-\xi_{3}]$. From Eq.(20), we
obtain $\tau$ as

\begin{equation}
\tau=
\left\{%
\begin{array}{ll}
    1+2\sqrt{\xi_{1}}s+\frac{2D}{\sqrt{\xi_{1}}} \\
    1-2\sqrt{\xi_{1}}s-\frac{2D}{\sqrt{\xi_{1}}} \\
    1+2\sqrt{\xi_{1}}s-\frac{2D}{\sqrt{\xi_{1}}} \\
    1-2\sqrt{\xi_{1}}s+\frac{2D}{\sqrt{\xi_{1}}} \\
\end{array}%
\right.
\end{equation}%
Imposing the condition $\tau^{'}\prec 0$, physical solutions are
given by two cases: \\
Case I: $k=-\xi_{2}+2D$,
$\pi=2\eta-1-\sqrt{\xi_{1}}s-\frac{D}{\sqrt{\xi_{1}}}$,
$\tau=1-2\sqrt{\xi_{1}}s-\frac{2D}{\sqrt{\xi_{1}}}$ \\
From Eq.(19) we obtain energy equation as

\begin{equation}
(2n+1)\sqrt{\xi_{1}}=-\xi_{2}+2D
\end{equation}%
Substituting $\xi_{1}$, $\xi_{2}$ and $D$ in Eq.(9), we solve
$\varepsilon$ as

\begin{equation}
\varepsilon=-\frac{B^{2}}{\lambda^{2}}(2A+1)^{2}\left[2n+1-2\sqrt{(2\eta-1)^{2}+
\frac{V_{0}}{\lambda^{2}}+2(\beta+1)-4A^{*}-4\eta(\eta-1)}\right]^{-2}
\end{equation}%
Using $\sigma(s)$ and $\pi(s)$ in Eqs.(16) and (26), we obtain the
corresponding wave functions $y(s)$ and $\phi(s)$. Then, from
Eq.(23) with

\begin{equation}
\rho(s)=s^{-\frac{2D}{\sqrt{\xi_{1}}}} e^{-2\sqrt{\xi_{1}}s}
\end{equation}%
we compute $y_{n}(s)$ form Eq.(22) as

\begin{equation}
y_{n}(s)=B_{n}L^{-\frac{2D}{\sqrt{\xi_{1}}}}_{n}(2\sqrt{\xi_{1}}s)
\end{equation}%
where $B_{n}=1/n!$. From Eq.(24), we solve

\begin{equation}
\phi(s)=s^{2\eta-1-\frac{D}{\sqrt{\xi_{1}}}}e^{-\sqrt{\xi_{1}}s}
\end{equation}%
Thus, total wave function becomes

\begin{equation}
\psi(s)=B_{n}s^{2\eta-1-\frac{D}{\sqrt{\xi_{1}}}}e^{-\sqrt{\xi_{1}}s}L^{-\frac{2D}{\sqrt{\xi_{1}}}}_{n}(2\sqrt{\xi_{1}}s)
\end{equation}%
Case II: $k=-\xi_{2}-2D$,
$\pi=2\eta-1-\sqrt{\xi_{1}}s+\frac{D}{\sqrt{\xi_{1}}}$,
$\tau=1-2\sqrt{\xi_{1}}s+\frac{2D}{\sqrt{\xi_{1}}}$. \\
From Eq.(19) we obtain energy equation as

\begin{equation}
2(n+1)\sqrt{\xi_{1}}=-\xi_{2}-2D
\end{equation}%
Substituting $\xi_{1}, \xi_{2}$ and $D$ in Eq.(35) we obtain

\begin{equation}
\varepsilon=-\frac{B^{2}}{\lambda^{2}}(2A+1)^{2}\left[2n+1+\sqrt{(2\eta-1)^{2}+
\frac{V_{0}}{\lambda^{2}}+2(\beta+1)-4A^{*}-4\eta(\eta-1)}\right]^{-2}
\end{equation}%
Using the same weight function defined in Eq.(32), we obtain

\begin{equation}
y_{n}(s)=B_{n}L^{\frac{2D}{\sqrt{\xi_{1}}}}_{n}(2\sqrt{\xi_{1}}s)
\end{equation}%
and also

\begin{equation}
\phi(s)=s^{2\eta-1+\frac{D}{\sqrt{\xi_{1}}}}e^{-\sqrt{\xi_{1}}s}
\end{equation}%
so total wave function becomes

\begin{equation}
\psi_{n}(s)=B_{n}s^{2\eta-1+\frac{D}{\sqrt{\xi_{1}}}}e^{-\sqrt{\xi_{1}}s}L^{\frac{2D}{\sqrt{\xi_{1}}}}_{n}(2\sqrt{\xi_{1}}s)
\end{equation}%
Eq.(34) and (35) are the general solutions of mass dependent
Schr\"{o}dinger equation which is given by Eq.(5) for the potential
relation introduced in Eq.(7).

\subsection{Solution of the Morse potential}

In this case, we aim to obtain the solutions for the potential
relation in Eq.(7) by reducing the mass dependent Schr\"{o}dinger
equation in Eq.(5) to a well-known Schr\"{o}dinger equation with a
Morse potential. The generalized Morse potential is

\begin{equation}
V(x)=V_{1}e^{-2\alpha^{*} x}-V_{2}e^{-\alpha^{*} x}
\end{equation}%
We write the SE for the potential in Eq. {40} by using a variable
transformation, $s=\sqrt{V_{1}}e^{-\alpha^{*} x}$ as

\begin{equation}
\frac{d^{2}\psi}{ds^{2}}+\frac{1}{s}\frac{d\psi}{ds}+
\frac{1}{s^{2}}\left[-\gamma^{*2}s^{2}+\gamma^{*2}\frac{V_{2}}{\sqrt{V_{1}}}s-4\varepsilon^{*2}\right]\psi=0
\end{equation}%
and

\begin{equation}
\varepsilon^{*2}=-\frac{mE^{*}}{2\hbar^{2}\alpha^{*2}}
\end{equation}%
and

\begin{equation}
\gamma^{*2}=\frac{2m}{\hbar^{2}\alpha^{*2}}
\end{equation}%
Comparing Eqs.(14) and (41), we define

\begin{eqnarray}
  \xi_{1} &=& \gamma^{*2} \\
  -\xi_{2} &=& \frac{V_{2}}{\sqrt{V_{1}}}\gamma^{*2} \\
  \xi_{3} &=& -4\varepsilon^{*2}
\end{eqnarray}

\begin{equation}
D^{2}=\xi_{1}[(2\eta-1)^{2}-\xi_{3}]
\end{equation}%
or

\begin{equation}
D=2\gamma^{*}\varepsilon^{*}
\end{equation}%
For the values of $\eta=\frac{1}{2}$, mass dependent Schr\"{o}dinger
equation in Eq.(14) turns into Eq.(41) which is not mass dependent
form for the Morse potential. From energy equation which is given by
Eq.(29),

\begin{equation}
(2n+1)\gamma^{*}=\frac{V_{2}}{\sqrt{V_{1}}}\gamma^{*2}+4\gamma^{*}\varepsilon^{*}
\end{equation}%
is obtained. Thus, we solve

\begin{equation}
\varepsilon^{*}=\frac{1}{4}\left[2n+1-\frac{V_{2}}{\sqrt{V_{1}}}\gamma^{*}\right]
\end{equation}%
Taking $\hbar=1$, we obtain the energy eigenvalues

\begin{equation}
E^{*}=-\frac{1}{4}\alpha^{*2}\left[2n+1-\frac{V_{2}}{\sqrt{V_{1}}}\gamma^{*}\right]^{2}
\end{equation}%
From Eq.(34), we obtain

\begin{equation}
\psi_{n}(s)=B_{n}s^{-2\varepsilon^{*}}e^{-\gamma^{*}s}L^{-4\varepsilon^{*}}_{n}(2\gamma^{*}s)
\end{equation}%
or one can re-write the wave function from Eq.(50)

\begin{equation}
\psi_{n}(s)=B_{n}s^{-\frac{1}{2}\left(-2n+1-\frac{V_{2}}{\sqrt{V_{1}}}\gamma^{*}\right)}e^{-\gamma^{*}s}
L^{\left(2n+1-\frac{V_{2}}{\sqrt{V_{1}}}\gamma^{*}\right)}_{n}(2\gamma^{*}s)
\end{equation}%
Eq.(53) which is the solutions of Eq.(41) is obtained by using
Eq.(14). If we consider these solutions, solutions of
Schr\"{o}dinger equation including mass dependent parameters  can be
found.  This case is given in below.

\section{Solutions within mass dependent parameters}

If we recall Eqs.(11-13, 44-46),
\begin{equation}
\xi_{1}=-\frac{\varepsilon}{\lambda^{2}}=\gamma^{*2}
\end{equation}%

\begin{equation}
\xi_{2}=-\frac{B(2A+1)}{\lambda^{2}}=-\frac{V_{2}}{\sqrt{V_{1}}}\gamma^{*2}
\end{equation}%

\begin{equation}
-\xi_{3}=\frac{V_{0}}{\lambda^{2}}+2(\beta+1)-4A^{*}+1=4\varepsilon^{*2}
\end{equation}%
and

\begin{equation}
\varepsilon=-\lambda^{2}\gamma^{*2}=-\frac{\sqrt{V_{1}}}{V_{2}}B(2A+1)
\end{equation}%
For the values of $\varepsilon=-B^{2}$, one can obtain $B$ as

\begin{equation}
B=\frac{\sqrt{V_{1}}}{V_{2}}(2A+1)
\end{equation}%
If we take $\lambda=\gamma^{*}=\frac{1}{\alpha^{*}}=1$, we obtain
$B=1$

\begin{eqnarray}
  \frac{V_{2}}{\sqrt{V_{1}}} &=& 2A+1
\end{eqnarray}
and from Eq. (52), the wave function becomes

\begin{equation}
\psi=B_{n}s^{-\sqrt{\frac{V_{0}}{\lambda^{2}}+2(\beta+1)-4A^{*}+1}}
e^{-\frac{1}{\lambda}\sqrt{-\varepsilon}s}L^{-2\sqrt{\frac{V_{0}}{\lambda^{2}}+2(\beta+1)-4A^{*}+1}}_{n}
\left(\frac{2}{\lambda}\sqrt{-\varepsilon}s\right)
\end{equation}%
From Eq.(52), relation of the wave function can be written as

\begin{equation}
\psi=B_{n}s^{-(n-A)}e^{-s}L^{-2(n-A)}_{n}(2s).
\end{equation}%
Using Eq.(36), $\varepsilon=-B^{2}$, and Eqs.(58, 59), we get

\begin{equation}
1=(2A+1)^{2}\left[2n+1-2\sqrt{\frac{V_{0}}{\lambda^{2}}+2(\beta+1)-4A^{*}+1}\right]^{-2}.
\end{equation}%
From Eq. (50), the energy is obtained. Substituting the expression
of $A^{*}$ in Eq.(9) and for the values of $\lambda=1$, $V_{0}$

\begin{equation}
V_{0}=(n-A)^{2}+4\alpha(\alpha+\beta+1)+2\beta+1
\end{equation}%
is obtained. The energy relation is obtained from Eqs.(42) and (59)
can be given below

\begin{equation}
E^{*}=-(n-A)^{2}.
\end{equation}%
If this expression is used in Eq.(63), $V_{0}$ has the following
form

\begin{equation}
V_{0}=-E^{*}+4\alpha(\alpha+\beta+1)+2\beta+1.
\end{equation}%

\section{Conclusions}

The effective mass Schr\"{o}dinger equation is solved for a given
potential. The Nikiforov-Uvarov method. is used to get energy
eigenvalues and the corresponding wave functions in a general form
by introducing a free parameter. By using this general form of the
solutions of the effective mass  Schr\"{o}dinger equation, we have
solved the effective mass Schr\"{o}dinger equation for Morse
potential transforming into a constant mass Schr\"{o}dinger
equation. We have shown that the effective mass Schr\"{o}dinger
equation can also be obtained starting from the constant mass case.

\section{Acknowledgements}
This research was partially supported by the Scientific and
Technological Research Council of Turkey.

\end{document}